\documentclass[aps,11pt]{revtex4}
\usepackage{amsfonts}
\usepackage[T2A]{fontenc}
\usepackage[cp1251]{inputenc}
\usepackage{amsmath}
\usepackage{amssymb}
\usepackage[russian,english]{babel}
\usepackage{graphicx}

\begin{document}

\title{On the stability of the coherent state of a two-level atom
Bose gas in the resonant laser field at zero temperature}

\author{L. A. Maksimov, A. V. Paraskevov}

\thanks{e-mail: paraskevov@kurm.polyn.kiae.su}

\affiliation{\\ Kurchatov Institute, Kurchatov Sq. 1, Moscow
123182, Russia}

\begin{abstract}

It is shown that a Bose gas of two-level atoms in the intense
resonant laser field at zero temperature is a mixture of two
condensates with a definite ratio of the densities. The criteria
of stability are found for the stationary states of such system
against the increment of the amplitudes of quasi Bogoliubov
elementary excitations. Besides the usual acoustic mode the gap
mode is shown to exist and the magnitude of the gap is
proportional to the laser field amplitude. The involvement of the
gas nonideality under definite conditions results in an
instability and decay of the condensates.

\end{abstract}

\maketitle PACS: 03.75.Kk, 03.75.Mn

\bigskip

\textbf{1. Introduction}

\bigskip

After the experimental realization of the Bose-Einstein
condensation in a gas \cite{1} and later in a binary mixture of
Bose gases \cite{2} in magnetic traps at ultralow temperatures the
adequate theoretical description of such systems is undoubtedly
actual. This is confirmed by a great number of theoretical papers
on the classification \cite{3} of all states of a Bose mixture
(within the Thomas-Fermi approximation) and its dynamics (longwave
collective excitations \cite{4}, metastable states \cite{5}, the
spatial separation of the mixture components \cite{6}). In most
papers the mixture is described by two Gross-Pitaevskii equations
and the concrete distinction between the mixture components is not
considered. The representation of the mixture components as atoms
in the ground and excited states is employed in the works on
studying the scattering \cite{7} and absorption \cite{8} of the
laser radiation by the Bose condensate. However, the question
about an existence of the stationary coherent state of the
interacting system, "mixture + laser field", has not been
considered yet. The present work is devoted to the theoretical
investigation of this problem. In particular, the equilibrium
ratio of the condensate densities is determined as a function of
the resonant laser field amplitude and proximity to the
saturation, the spectrum of quasi Bogoliubov elementary
excitations is found, and the criteria of stability in the system
of two condensates in the laser field are determined with the
involvement of the recoil momentum and nonideality of a gas.

Let us consider a weakly non-ideal Bose gas of two-level atoms in
the resonant laser field of the high intensity at zero
temperature. The monochromatic laser field, acting on the Bose
condensate of atoms in the ground state (for simplicity, we
neglect noncondensate particles), converts some fraction of the
atoms into the dipole-excited state so that each excited atom
moves with the same velocity governed by the recoil momentum. If
the laser field intensity is so high that one can neglect the
spontaneous decay of excited atoms with respect to the induced
emission, then, during the small time as compared with the time
between collisions of excited atoms with quasiparticles of the
condensate of unexcited atoms, the macroscopically large number of
dipole-excited atoms moving with the same velocity is formed and
the second Bose condensate appears. Below we do not treat the
process of the formation of this condensate and assume that the
state with two condensate has been already formed, taking as a
condition that the natural linewidth of dipole transition is
$\gamma\approx0$. The criterion of a smallness for $\gamma$ will
be given in Sec. 2. At last, from the viewpoint of maximal
simplicity of the effect treatment, the atomic system is
considered as spatially homogeneous.

\bigskip

\textbf{2. The ground state of two-level gas in the laser field}

\bigskip

The Hamiltonian for the system consisting of $N_{a}$ unexcited
atoms and  $N_{b}$ dipole-excited atoms in the laser field
($V=1,\hbar=1$) reads
\begin{equation}%
\begin{array}
[c]{c}%
H_{tot}=H_{gas}+H_{int}+H_{ph},\ H_{gas}=H_{a}+H_{b}+H_{ab},\\
H_{a}=\sum\limits_{p}(\frac{p^{2}}{2m})\hat{a}_{p}^{+}\hat{a}_{p}%
+\sum\limits_{p_{1}+p_{2}=p_{3}+p_{4}}\frac{1}{2}U_{aa}\hat{a}_{p_{1}}^{+}%
\hat{a}_{p_{2}}^{+}\hat{a}_{p_{3}}\hat{a}_{p_{4}},\\
H_{b}=\sum\limits_{p}(\frac{p^{2}}{2m}+\omega_{0})\hat{b}_{p}^{+}\hat{b}%
_{p}+\sum\limits_{p_{1}+p_{2}=p_{3}+p_{4}}\frac{1}{2}U_{bb}\hat{b}_{p_{1}}%
^{+}\hat{b}_{p_{2}}^{+}\hat{b}_{p_{3}}\hat{b}_{p_{4}},\\
H_{ab}=\sum\limits_{p_{1}+p_{2}=p_{3}+p_{4}}U_{ab}\hat{a}_{p_{1}}^{+}\hat
{b}_{p_{2}}^{+}\hat{b}_{p_{3}}\hat{a}_{p_{4}}.
\end{array}
\label{1}%
\end{equation}
Here $\hat{a}_{p}$ and $\hat{b}_{p}$ are the atom annihilation
operators with momentum $p$ in the ground and dipole-excited
states, respectively, and $U_{ik}=4\pi a_{ik}/m$ are the
parameters of the pair interaction expressed in terms of the
corresponding scattering lengths. To exclude the problem of
collapse, we suppose $U_{ik}>0$. The interaction of the gas with a
single-mode laser field is described by the Hamiltonian
\begin{equation}
H_{int}=\sum_{p}(g_{k}\hat{c}_{k}^{+}\hat{a}_{p}^{+}\hat{b}_{p+k}+g_{k}^{\ast
}\hat{b}_{p+k}^{+}\hat{a}_{p}\hat{c}_{k}), \label{2}%
\end{equation}
where $\hat{c}_{k}$ and $\hat{c}_{k}^{+}$ are the absorption and
production operators of a photon with wave vector $\vec{k}$ and
frequency $\omega_{k}$. For simplicity, the laser radiation is
assumed to be linearly polarized in the direction of unit
polarization vector $\vec{e}_{k}$. The atom-field coupling
constant has a usual form
\[
\left\vert g_{k}\right\vert ^{2}=2\pi\omega_{k}\left\vert
(\vec{e}_{k},\vec
{d})\right\vert ^{2}%
\]
and is related to the natural linewidth of the excited level for a
single atom via dipole matrix element $d$,
$\gamma=\frac{4}{3}k_{0}^{3}d^{2}$ where
$k_{0}=\frac{\omega_{0}}{c}$. To simplify the notation, the factor
$\exp(i\vec{k}\vec{r})$ in the matrix element of the interaction
between a single atom and laser field is omitted.
Lastly, the
Hamiltonian of free photons is given by
\[
H_{ph}=\omega_{k}\hat{c}_{k}^{+}\hat{c}_{k}.
\]
The dynamics of the system is described by the equations
\begin{equation}%
\begin{array}
[c]{c}%
i\frac{\partial\hat{a}_{p}}{\partial t}=\frac{p^{2}}{2m}\hat{a}_{p}%
+\sum\limits_{p+p_{2}=p_{3}+p_{4}}U_{aa}\hat{a}_{p_{2}}^{+}\hat{a}_{p_{3}}%
\hat{a}_{p_{4}}+\sum\limits_{p_{1}+p=p_{3}+p_{4}}U_{ab}\hat{b}_{p_{1}}^{+}%
\hat{b}_{p_{3}}\hat{a}_{p_{4}}+g_{k}\hat{c}_{k}^{+}\hat{b}_{p+k},\\
i\frac{\partial\hat{b}_{p}}{\partial
t}=(\frac{p^{2}}{2m}+\omega_{0})\hat
{b}_{p}+\sum\limits_{p+p_{2}=p_{3}+p_{4}}U_{bb}\hat{b}_{p_{2}}^{+}\hat
{b}_{p_{3}}\hat{b}_{p_{4}}+\sum\limits_{p_{1}+p=p_{3}+p_{4}}U_{ab}\hat
{a}_{p_{1}}^{+}\hat{b}_{p_{3}}\hat{a}_{p_{4}}+g_{k}^{\ast}\hat{a}_{p-k}\hat
{c}_{k},\\ i\frac{\partial\hat{c}_{k}}{\partial
t}=\omega_{k}\hat{c}_{k}+\sum
\limits_{p}g_{k}\hat{a}_{p}^{+}\hat{b}_{p+k}.
\end{array}
\label{03}%
\end{equation}
Let us consider the system which consists of three coherent
subsystems, namely, two condensates produced by atoms in the
ground and dipole-excited states and the laser field. Let us find
the ground state of this system, neglecting noncondensate
particles. In what follows, it is more convenient to deal with
usual complex quantities rather than with operators. Therefore the
use of the representation of coherent states is appropriate. Let
us determine a coherent state as
\[
\left\vert a\right\rangle
=\exp(\sum\limits_{p}a_{p}\hat{a}_{p}^{+})\left\vert
0\right\rangle ,
\]
which can be considered as a direct product of all eigenvectors
$\left\vert a_{p}\right\rangle $ so that $\hat{a}_{p}\left\vert
a_{p}\right\rangle =a_{p}\left\vert a_{p}\right\rangle
,\left\langle a_{p}\right\vert \hat{a}_{p}^{+}=\left\langle
a_{p}\right\vert a_{p}^{\ast}$ and concerning the states with
different momenta. Vector ($\left\vert 0\right\rangle $ denotes a
direct product of all empty atom and photon states. For
dipole-excited atoms and photons, the coherent states are
determined similarly:
\[
\left\vert b\right\rangle
=\exp(\sum\limits_{p}b_{p}\hat{b}_{p}^{+})\left\vert
0\right\rangle ,\text{ \ }\left\vert c\right\rangle =\exp(c_{k}\hat{c}_{k}%
^{+})\left\vert 0\right\rangle .
\]
Taking the recoil into account, condensates of atoms in the ground
and dipole-excited states should move relatively each other. Let
us denote the momenta of atoms belonging to these condensates
$\vec{p}_{a}$ and $\vec{p}_{b}=\vec{p}_{a}+\vec{k}$, respectively.
If the gas is at rest as a whole, $\vec{p}_{a}=-\vec{k}
N_{b}/(N_{a}+N_{b})$. Postmultiplying each of equations (\ref{03})
by vector $\left\vert a\right\rangle \left\vert b\right\rangle
\left\vert c\right\rangle $, we obtain the set of three equations
for the complex fields $a_{p_{a}},\, b_{p_{b}},c_{k}$:
\begin{equation}
\begin{array}
[c]{c}%
i\frac{\partial a_{p_{a}}}{\partial t}=(\frac{p_{a}^{2}}{2m}+U_{aa}%
N_{a}+U_{ab}N_{b})a_{p_{a}}+g_{k}c_{k}^{\ast}b_{p_{b}},\\
i\frac{\partial b_{p_{b}}}{\partial t}=(\frac{p_{b}^{2}}{2m}+\omega_{0}%
+U_{bb}N_{b}+U_{ab}N_{a})b_{p_{b}}+g_{k}^{\ast}a_{p_{a}}c_{k},\\
i\frac{\partial c_{k}}{\partial
t}=\omega_{k}c_{k}+g_{k}a_{p_{a}}^{\ast }b_{p_{b}}.
\end{array}
\label{04}
\end{equation}
Hereafter we consider a gas in the condensate state. Therefore
$N_{a}$ and $N_{b}$ are macroscopically large quantities, i.e.,
$N_{a},\, N_{b}\gg1$.  We seek a stationary solution in the form
\begin{equation}%
\begin{array}
[c]{c}%
a_{p_{a}}=ae^{-i\varepsilon_{a}t},\ b_{p_{b}}=be^{-i\varepsilon_{b}t}%
,\ c_{k}=ce^{-i\omega t}.
\end{array}
\label{05}%
\end{equation}
Note that a stationary solution (\ref{05}) takes place only under
condition of the exact resonance between the renormalized laser
frequency and the difference in energies of the excited and ground
states of an atom
\begin{equation}
\omega=\varepsilon_{b}-\varepsilon_{a}. \label{006}%
\end{equation}
From (\ref{04}) we get%
\begin{equation}%
\begin{array}
[c]{c}%
\varepsilon_{a}a=\Omega_{a}a+g_{k}c^{\ast}b,\\
\varepsilon_{b}b=\Omega_{b}b+g_{k}^{\ast}ac,\\ \left(
\varepsilon_{b}-\varepsilon_{a}\right)
c=\omega_{k}c+g_{k}a^{\ast}b.
\end{array}
\label{002}%
\end{equation}
Here the following notations are introduced
\[
\Omega_{a}=\frac{\vec{p}_{a}^{2}}{2m}+U_{aa}N_{a}+U_{ab}N_{b},\
\Omega
_{b}=\frac{\vec{p}_{b}^{2}}{2m}+\omega_{0}+U_{ab}N_{a}+U_{bb}N_{b},\text{
}N_{a}=\left\vert a\right\vert ^{2},\ N_{b}=\left\vert
b\right\vert ^{2}.
\]
Quantities  $\varepsilon_{a}$, $\varepsilon_{b}$ and $N_{a}$ (or
$N_{b}$ since we consider that the given total gas density is
$N=N_{a}+N_{b}=const$) are unknown.  The third equation in
(\ref{002}) determines the renormalization of the laser frequency
due to dipole transitions. This renormalization can be neglected
if the laser field intensity is sufficiently high (formally, this
means that $H_{ph}\gg H_{int}$, implying matrix elements) so that
\begin{equation}
N_{c}=\left\vert c\right\vert ^{2}\gg\frac{\left\vert
g_{k}\right\vert
^{2}N_{a}N_{b}}{\omega_{k}^{2}}. \label{909}%
\end{equation}
The condition of the smallness of the natural linewidth results
directly from this inequality
\begin{equation}
\gamma\ll\omega_{k}\frac{N_{c}k_{0}^{3}}{N_{a}N_{b}}. \label{1011}%
\end{equation}
Inequality (\ref{909}) is worthwhile to compare with the initially
implied condition of the sufficiently high field intensity in
order to neglect spontaneous decays with respect to the induced
transitions. In the terms of the notations introduced this
condition is given by
\begin{equation}
N_{c}\gg\frac{2}{\pi}k^{3}. \label{333}%
\end{equation}
Note that, if one takes resonance approximation ($\left\vert
\omega_{k} -\omega_{0}\right\vert \ll\omega_{k},\omega_{0}$) into
account, inequality (\ref{333}) yields the condition known in the
laser physics for saturation of the population of excited state,
when an atom with the equal probability is in the ground and
excited states so that $N_{a}=N_{b}=N/2$. The inequality is
expressed quantitatively as
\begin{equation}
\left\vert \omega_{k}-\omega_{0}\right\vert
\sim\gamma\ll\left\vert
A\right\vert , \label{444}%
\end{equation}
where $\left\vert A\right\vert =\left\vert g_{k}\right\vert
\sqrt{N_{c}}$ is the Rabi frequency in the conventional
terminology. From (\ref{444}) we obtain
\begin{equation}
\gamma\ll\omega_{k}\frac{N_{c}}{k_{0}^{3}}. \label{555}%
\end{equation}
From the comparison of conditions (\ref{1011}) and (\ref{555}) one
can see that they are equivalent to each other if $k_{0}^{3}\geq
N$, i.e., when the volume per one atom is larger than
$\lambda_{0}^{3}$. According to (\ref{333}), this means
qualitatively that a single atom is fallen at many photons. Note,
however, that condition (\ref{444}) looses its meaning if the
renormalization of the transition frequency  is larger than the
natural linewidth for a single atom, $\left\vert \varepsilon
_{b}-\varepsilon_{a}-\omega_{0}\right\vert >\gamma$. In this case,
it seems that $\gamma$ should be regarded as a total natural
linewidth with respect to the collisional broadening. The
calculation of the corresponding contribution is made in \cite{8}.

In what follows, we assume that inequality (\ref{1011}) is
fulfilled and $\omega\simeq\omega_{k}$. Denoting
$\varepsilon=\varepsilon_{a}$ and re-denoting
$\Omega_{b}\longrightarrow\Omega_{b}-\omega$,  we have from the
equations (\ref{002})
\[
\left(  \varepsilon-\Omega_{a}\right)  \left(
\varepsilon-\Omega_{b}\right) =\left\vert A\right\vert ^{2}.
\]
Hence
\begin{equation}
\varepsilon=\frac{1}{2}\left[
(\Omega_{a}+\Omega_{b})+\eta\sqrt{\delta
\omega^{2}+4\left\vert A\right\vert ^{2}}\right]  ,\eta=\pm1. \label{130}%
\end{equation}
Here the quantity
\[
\delta\omega=\Omega_{a}-\Omega_{b}=\omega-\omega_{0}+\frac{\vec{k}^{2}}%
{2m}\left(  \frac{N_{b}-N_{a}}{N_{a}+N_{b}}\right)  +U_{aa}N_{a}-U_{bb}%
N_{b}\ +U_{ab}(N_{b}-N_{a})
\]
plays a role of the deviation from the full saturation. The
relation between the amplitudes of the condensates is determined
from (\ref{002})
\[
b=\left[  -\frac{\delta\omega}{2\left\vert A\right\vert
}+\eta\sqrt{\left( \frac{\delta\omega}{2\left\vert A\right\vert
}\right)  ^{2}+1}\right]  a.
\]
It follows from above that the system in the laser field is in one
of the states with the fixed  ratio of the condensate densities
(at the given sign of product $\eta\delta\omega$)
\begin{equation}
N_{b}=\left[  -\eta\frac{\delta\omega}{2\left\vert A\right\vert }%
+\sqrt{\left(  \frac{\delta\omega}{2\left\vert A\right\vert }\right)  ^{2}%
+1}\right]  ^{2}N_{a}. \label{6}%
\end{equation}
When $\delta\omega=0$, the resonance field frequency equals
$\omega=\omega _{0}+\frac{1}{2}(U_{bb}-U_{aa})N$. Thus, if
$U_{aa}\simeq U_{bb}$, it is natural to take difference
$\omega-\omega_{0}$ as an external parameter characterizing the
deviation from the saturation.  Let us introduce the dimensionless
variables
\begin{equation}
x=\frac{\omega-\omega_{0}}{2\left\vert A\right\vert },\ y=\frac{N_{b}-N_{a}%
}{N},\ z\simeq\frac{1}{2\left\vert A\right\vert }[\frac{k^{2}}{2m}%
+(U_{ab}-U_{aa})N] \label{88}%
\end{equation}
and write equation (\ref{6}) in the form
\begin{equation}
1+y=\left(  -\eta\left(  x+zy\right)  +\sqrt{\left(  x+zy\right)  ^{2}%
+1}\right)  ^{2}(1-y). \label{45}%
\end{equation}
Equation (\ref{45}) is a quartic one with respect to $y\left(
x\right)  $. The solution relative to the inverse function
$x\left(  y\right)  $ has a simple form
\begin{equation}
x=-(z+\frac{\eta}{\sqrt{1-y^{2}}})y. \label{48}%
\end{equation}
It follows from the above expression that there exists a symmetry
with respect to the simultaneous substitution $x\longrightarrow-x$
and $y\longrightarrow-y$. This results from the symmetry of the
Hamiltonian of the system with regard to substitution
$\hat{a}\leftrightarrow \hat{b}$. In its turn, such symmetry
between "up" and "down" is a result of the assumption about the
smallness of $\gamma$. Note that there is no symmetry in the
solutions with regard to the variation of the sign $\eta$. For the
branch $\eta=+1$, quantity $y\left( x\right)  $ is a one-to-one
function of $x$ and, for $\eta=-1$, function $y\left(  x\right)$
at $z>1$ within the interval \[
\left\vert x\right\vert <x_{\lim}=\left(  z^{2/3}-1\right)  ^{3/2}%
\]
takes four values (see Fig.1). Accordingly, for the branch
$\eta=-1$ at $z>1$ there exists limiting values $y(\pm
x_{\lim})=\mp\sqrt{1-z^{-2/3}} $ with an anomalous S-shape of the
line in comparison with the case $z<1$.

For $x=0$, both condensates have either the same density, $y=0$,
for any sign or
\begin{equation}
y=\pm\sqrt{1-z^{-2}}, \label{39}%
\end{equation}
at $\eta=-1$. Thus, the satellite states are possible if the
coupling between atoms and field is sufficiently small compared
with the recoil energy $\varepsilon_{r} =\frac{k^{2}}{2m}$ ($z>1$,
being $z\approx\frac{\varepsilon_{r}}{2\left\vert A\right\vert }$
for $U_{ab}\approx U_{aa},U_{bb}$). If the coupling between atoms
and field is strong, $z<1$, there are solutions only in the
vicinity of the centre ($y=-\frac{1}{z+\eta}x$) at $\left\vert
x\right\vert \ll1$ (i.e., in the case $\left\vert
\varepsilon_{b}-\varepsilon_{a}-\omega_{0}\right\vert
\leq\gamma\ll\left\vert A\right\vert $).

So, it is seen from the analysis of function  $x\left( y\right)$
that (i) under the condition of saturation $\left\vert
x\right\vert \ll1$ there exist four stationary states, where in
addition to the usual solutions  $ y\approx0$  there are two
satellite states in which populations of the ground and excited
states are essentially different. (ii) Beyond the region of
saturation at $\left\vert x\right\vert
>x_{\lim}$ there are two stationary states, and at $\left\vert x\right\vert <x_{\lim}$
there are four ones again. It is also interesting to trace the
behavior $y(z)$ at $x=0$. If $z>1$, bifurcation, associated with a
possibility of satellite states, appears in the plot $y(z)$ in
addition to $y=0$ (see Fig.2).

\bigskip

\textbf{3. Combined oscillations of the condensates in the laser
field}

\bigskip

The stationary state obtained is unstable if the spectrum of
elementary excitations of the system in this state has an
imaginary part. To clarify a role of the interaction between
atoms, we start from the case of an ideal gas. In addition, we
will consider the question about the stability in the case of full
saturation $\omega=\omega_{0},N_{a}=N_{b}=N/2$ and in the case of
a satellite state.

First, we find a set of equations determining spectrum of
single-particle excitations of the condensates. Let us write
equations (\ref{03}) for the complex fields in the linear
approximation in the amplitudes of noncondensate particles in the
following representation (hereafter $\vec{q}$ is a quasiparticle
momentum):
\begin{align*}
a_{p}  &  =\bar{a}_{p}\exp\left(  -i\varepsilon t\right)  ,\text{
\ }\vec {p}=\vec{p}_{a}\pm\vec{q},\\ b_{p}  &
=\bar{b}_{p}\exp\left(  -i\varepsilon t-i\omega t\right)  ,\text{
\ }\vec{p}=\vec{p}_{b}\pm\vec{q},
\end{align*}
in which the system (\ref{03}) takes a form of a closed system
with constant coefficients (vector notations are omitted):
\begin{equation}%
\begin{array}
[c]{c}%
i\frac{\partial\bar{a}_{p_{a}+q}}{\partial t}=(\frac{\left(
p_{a}+q\right)
^{2}}{2m}-\varepsilon+2\mu_{aa}+U_{ab}N_{b})\bar{a}_{p_{a}+q}+\mu_{aa}\bar
{a}_{p_{a}-q}^{+}+(\mu_{ab}+A)\bar{b}_{p_{b}+q}+\mu_{ab}\bar{b}_{p_{b}-q}%
^{+},\\ -i\frac{\partial\bar{a}_{p_{a}-q}^{+}}{\partial
t}=(\frac{\left(
p_{a}-q\right)  ^{2}}{2m}-\varepsilon+2\mu_{aa}+U_{ab}N_{b})\bar{a}_{p_{a}%
-q}^{+}+\mu_{aa}\bar{a}_{p_{a}+q}+\mu_{ab}\bar{b}_{p_{b}+q}+(\mu_{ab}%
+A)\bar{b}_{p_{b}-q}^{+},\\
i\frac{\partial\bar{b}_{p_{b}+q}}{\partial t}=(\frac{\left(
p_{b}+q\right)
^{2}}{2m}-\varepsilon-\left(  \omega-\omega_{0}\right)  +2\mu_{bb}+U_{ab}%
N_{a})\bar{b}_{p_{b}+q}+\mu_{bb}\bar{b}_{p_{b}-q}^{+}+(\mu_{ab}+A)\bar
{a}_{p_{a}+q}+\mu_{ab}\bar{a}_{p_{a}-q}^{+},\\
-i\frac{\partial\bar{b}_{p_{b}-q}^{+}}{\partial t}=(\frac{(p_{b}-q)^{2}}%
{2m}-\varepsilon-\left(  \omega-\omega_{0}\right)  +2\mu_{bb}+U_{ab}N_{a}%
)\bar{b}_{p_{b}-q}^{+}+\mu_{bb}\bar{b}_{p_{b}+q}+\mu_{ab}\bar{a}_{p_{a}%
+q}+(\mu_{ab}+A)\bar{a}_{p_{a}-q}^{+},
\end{array}
\label{99}%
\end{equation}
where the notations $\mu_{ik}=U_{ik}\sqrt{N_{i}N_{k}}$ and
$A=g_{k}^{\ast}c$ are used. The phase of laser field is chosen so
that quantity $A$ is real and positive.

In the case of an ideal gas, when all $U_{ik}=0$, the equation for
eigenvalues of the system (\ref{99}),  with involving (\ref{130})
and (\ref{88}), yields
\begin{align*}
E_{1,2}  &
=-\frac{\vec{k}\vec{q}}{2m}y+\frac{q^{2}}{2m}-\frac{1}{2}\eta
\sqrt{\delta\omega^{2}+4\left\vert A\right\vert ^{2}}\pm\frac{1}{2}%
\sqrt{\left(  \delta\omega+\frac{\vec{k}\vec{q}}{m}\right)
^{2}+4\left\vert A\right\vert ^{2}},\\ E_{3,4}  &
=-\frac{\vec{k}\vec{q}}{2m}y-\frac{q^{2}}{2m}+\frac{1}{2}\eta
\sqrt{\delta\omega^{2}+4\left\vert A\right\vert ^{2}}\pm\frac{1}{2}%
\sqrt{\left(  \delta\omega-\frac{\vec{k}\vec{q}}{m}\right)
^{2}+4\left\vert A\right\vert ^{2}}.
\end{align*}
Here $\delta\omega=\omega-\omega_{0}+\frac{k^{2}}{2m}y$. For the
full saturation ($\delta\omega=0,y=0$), we have
\[
E_{1,2}=\left(  \frac{q^{2}}{2m}-\eta\left\vert A\right\vert
\right)  \pm \sqrt{\left(  \frac{\vec{k}\vec{q}}{2m}\right)
^{2}+\left\vert A\right\vert ^{2}},\text{ \
}E_{3,4}=-\frac{q^{2}}{2m}+\eta\left\vert A\right\vert \pm
\sqrt{\left(  \frac{\vec{k}\vec{q}}{2m}\right)  ^{2}+\left\vert
A\right\vert ^{2}}.
\]
Hence, for the given $\eta$ there is a gap $2\left\vert
A\right\vert $ for two modes in the excitation spectrum at $q=0$.
The gapless modes at $A=0$ and $q\longrightarrow0$ yield a sound
dispersion law
\[
E(q)\sim c_{s}q,
\]
where $c_{s}=\frac{\left\vert \vec{k}\vec{q}\right\vert }{2mq}$ is
the sound velocity. Far from the saturation ($\left\vert
x\right\vert \gg1,\left\vert y\right\vert \simeq1$) in (\ref{88})
for the given $\eta$ the two branches have a gap of about
$\left\vert \omega-\omega_{0}\right\vert $ and the others are
gapless as before.

Thus, in the approximation of an ideal gas the stationary
condensate state is always stable in the sense mentioned above.

In the case of nonideal gas one can find an analytical expression
for the spectrum of the system (\ref{99}) under full saturation,
assuming that $U_{aa}=U_{bb}$:
\begin{align}
E_{\pm}^{2}  &  =\left(  P^{2}-\mu^{2}\right)
+A(A+2\mu_{ab})+Q^{2}\pm
2\sqrt{(P^{2}-\mu^{2})Q^{2}+\left(  AP+\mu_{ab}(P-\mu)\right)  ^{2}%
},\label{1000}\\
P  &  =\frac{q^{2}}{2m}-\eta A+\mu,\text{ \ }Q=\frac{\vec{k}\vec{q}}%
{2m},\ \text{\ }\mu=\frac{1}{2}U_{aa}N,\text{ \ }\mu_{ab}=\frac{1}{2}%
U_{ab}N.\nonumber
\end{align}
First of all, we consider the long wave limit $q\longrightarrow0$
and $E_{\pm }^{2}(q)\approx E_{\pm}^{2}(0)+\frac{q^{2}}{2m}\Delta
E$ where
\begin{equation}%
\begin{array}
[c]{c}%
E_{\pm}^{2}(0)=\xi^{2}-\mu^{2}+A\left(  A+2\mu_{ab}\right)
\pm2A\left\vert \xi-\eta\mu_{ab}\right\vert ,\text{ \
}\xi=\mu-A\eta,\\ \Delta
E=2\xi+\varepsilon_{r}\cos^{2}\theta\pm\left(  \frac{\left(  \xi
^{2}-\mu^{2}\right)  }{A\left\vert \xi-\eta\mu_{ab}\right\vert }%
\varepsilon_{r}\cos^{2}\theta+2A\left(  A+\mu_{ab}\right)
sign\left( \xi-\eta\mu_{ab}\right)  \right)  .
\end{array}
\label{1234}%
\end{equation}
Here $\theta$ is the angle between vectors $\vec{k}$ and $\vec{q}$
. The critical value, in the sense of stability, is $E_{-}^{2}$.
In the case when the argument of modulus in (\ref{1234}) is
positive $\mu>\eta\left( \mu_{ab}+A\right)  $ we obtain
\begin{align*}
E_{-}^{2}(0) &  =2A(A-\mu+\mu_{ab})(1+\eta),\\ \Delta E &
=2\left(  \mu-\mu_{ab}-A\left(  1+\eta\right)  \right)
+\varepsilon_{r}\cos^{2}\theta\left[  1-\frac{\left(  A-2\eta\mu\right)  }%
{\mu-\eta\left(  \mu_{ab}+A\right)  }\right]  .
\end{align*}
It is seen that there is an instability in the state $\eta=1$ at
$\mu
>\mu_{ab}+A$ and in the state $\eta=-1$
\[
\Delta E=2\left(  \mu-\mu_{ab}\right)  \left(  1-\frac{1}{2}\frac
{\varepsilon_{r}\cos^{2}\theta}{\mu+\mu_{ab}+A}\right)  ,
\]
at $\mu<\mu_{ab}$. This coincides with the known condition
\cite{6} of instability of a binary Bose-gas mixture against
spatial separation of its components. The instability at
$\varepsilon_{r}\cos^{2}\theta>2\left(  \mu+\mu_{ab}+A\right)$
means that for the strong recoil the instability develops at first
in the direction of the recoil momentum. The corresponding plot of
the spectrum (\ref{1000}) at $\cos ^{2}\theta=1$ is given in
Fig.3. The opposite case $\mu<\eta\left(  \mu_{ab}+A\right)  $ has
a sense only for $\eta=1$ and does not result in an instability
\[
E_{-}^{2}(0)=0,\text{ \ }\Delta E=2\left(  \mu+\mu_{ab}\right)
\left(
1+\frac{1}{2}\frac{\varepsilon_{r}\cos^{2}\theta}{A+\mu_{ab}-\mu}\right)
>0.
\]
Let us now consider the  case of transverse excitations with
respect to vector $\vec{k}$ when $Q=0$
\begin{equation}
E_{\pm}^{2}=\delta^{2}+2\mu\delta+A(A+2\mu_{ab})\pm2\left\vert
A\left(
\delta+\mu\right)  +\mu_{ab}\delta\right\vert ,\text{ \ }\delta=\frac{q^{2}%
}{2m}-\eta A.\label{103}%
\end{equation}
For $E_{-}^{2}$ at $\left(  A+\mu_{ab}\right)  \left(
\frac{q^{2}}{2m}-\eta A\right)  +A\mu>0$, we get
\[
E_{-}^{2}=\left(  \frac{q^{2}}{2m}-A\left(  1+\eta\right)  \right)
\left( \frac{q^{2}}{2m}-A\left(  1+\eta\right)  +2\left(
\mu-\mu_{ab}\right) \right)  .
\]
It is seen that in the state $\eta=-1$ an instability appears for
the condition obtained above $\mu<\mu_{ab}$. In the state $\eta=1$
there is a region of instability $2A-2\left( \mu-\mu_{ab}\right)
<\frac{q^{2}}{2m}<2A$ with the centre at the point
$\frac{q_{\min}^{2}}{2m}=2A-\left(  \mu-\mu_{ab}\right)  $ where $E_{-}%
^{2}(q_{\min})=-\left(  \mu-\mu_{ab}\right)  ^{2}$. In addition,
for $q=0$ in this state there appears an instability if
$\mu>A+\mu_{ab}$ (however, it is always
$E_{-}^{2}(q_{\min})<E_{-}^{2}(0)$). The typical plot of the
spectrum for this case (\ref{1000}) is given in Fig.4.

In the case $\left(  A+\mu_{ab}\right)  \left(
\frac{q^{2}}{2m}-\eta A\right) +A\mu<0$ we have
\[
E_{-}^{2}=\left(  \frac{q^{2}}{2m}+A\left(  1-\eta\right)  \right)
\left( \frac{q^{2}}{2m}+A\left(  1-\eta\right)  +2\left(
\mu+\mu_{ab}\right) \right)  ,
\]
whence it follows that the both states $\eta=\pm1$ are stable.

Finally, let us consider the stability of satellites at
$U_{aa}=U_{bb}$. Far from the saturation $\left(  \left\vert
\omega-\omega_{0}\right\vert \gg A,\text{ }\mu,\text{
}\mu_{ab}\right) $ one can neglect the nonideality of the gas and,
as we convinced above, the condensates are stable. In the region
of saturation $\left\vert x\right\vert \ll1$, the satellites exist
only in the state with $\eta=-1$. For simplicity, we restrict
ourselves with the case when the density of one of condensates is
much less than that of the other. For definiteness, $N_{b}\ll
N_{a}$ $\left( y\simeq-1\right)  $. It turns out that one can
derive an analytical expression for the spectrum of elementary
excitations in the approximation of heavy atoms and strong
atom-atom coupling compared with the recoil, putting formally
$k\ll q$. Note that in these conditions it is not allowed to
employ the long wave limit $q\longrightarrow0$.
\begin{align}
E_{\pm}^{2} &  =\frac{1}{2}\left(
W^{2}+V^{2}+2A^{2}-\mu^{2}\pm\sqrt{\left(
W^{2}-V^{2}-\mu^{2}\right)  ^{2}+4A^{2}\left(  \left(  W+V\right)  ^{2}%
-\mu^{2}\right)  }\right)  ,\label{199}\\ W &
=\frac{q^{2}}{2m}+\frac{1}{2}\left(  3\mu-\mu_{ab}\right)  +\sqrt
{\frac{1}{4}\left(  \mu-\mu_{ab}\right)  ^{2}+A^{2}},\nonumber\\ V
&  =\frac{q^{2}}{2m}-\frac{1}{2}\left(  \mu-\mu_{ab}\right)
+\sqrt{\frac {1}{4}\left(  \mu-\mu_{ab}\right)
^{2}+A^{2}},\nonumber
\end{align}
where $\mu=U_{aa}N,$ $\mu_{ab}=U_{ab}N$. Rewriting expression
(\ref{199}) in the form
\[
E_{\pm}^{2} =\frac{1}{2}\left(
W^{2}+V^{2}+2A^{2}-\mu^{2}\pm\sqrt{\left(
W^{2}+V^{2}+2A^{2}-\mu^{2}\right)  ^{2}-4\left(  A^{2}-WV\right)  ^{2}%
+4\mu^{2}V^{2}}\right)  ,
\]
one can straightforwardly indicate the region of possible negative
values $E_{-}^{2}$,
\[
-4\left(  A^{2}-WV\right)  ^{2}+4\mu^{2}V^{2}>0.
\]
Hence
\begin{equation}
\mu V>\left(  WV-A^{2}\right)  .\label{200}%
\end{equation}
Note that the condition (\ref{200}) never holds true at
$\mu\simeq\mu_{ab}$ and the satellite state is stable.

\bigskip

\textbf{4. Conclusion}

\bigskip

The main result of the work is the determination of a stationary
solution of the equations of motion for the united coherent system
of two-level atom Bose condensates and laser field of high
intensity. In addition, the dependence of the condensate densities
on the Rabi frequency and deviation from the full saturation are
found. The analysis is performed for the stability of the system
against appearance of the imaginary part in the spectrum of
elementary excitations. This results in an exponential increase of
oscillations, in the heating of the system and decay of the
condensates, or in the transition of the system into one of the
stable states. The question on evolution of the system after the
decay of an unstable stationary state is a subject of separate
study.

The condition of applicability of the theory in temperature is
given by
\[
T\ll T_{c}\sim\frac{\hbar^{2}}{m}n^{2/3},\text{ \ }n=\min\left(  N_{a}%
,N_{b}\right)  .
\]
Note that in the satellite state, when the condensate densities
differ significantly, the condensate of the lower density at
$T>T_{c}$ converts into noncondensate particles which are
additional with respect to the condensate of higher density. The
question whether this condensate will be destroyed with adding
extrinsic noncondensate particles represents an independent
interest.

The authors are grateful to Yu. Kagan and S. N. Burmistrov for
valuable comments. The work is supported by the grants of Russian
Foundation of Basic Research and INTAS.

\newpage

\textbf{Figure Captions}

\bigskip

Fig.1. The dependence of the relative difference in populations,
$y$, versus deviation $x$ from the full saturation.

\bigskip

Fig.2. The dependence of the relative difference in populations,
$y$, versus parameter$\ z$ (see (\ref{88})).

\bigskip

Fig.3. The example of unstable spectrum in the long wave limit for
the state $\eta=-1$ under full saturation at $\varepsilon
_{r}>2\left( \mu +\mu _{ab}+A\right) =0.43\varepsilon _{r}$, and
$\mu :\mu _{ab}:A=1:0.5:20$. The scale on the vertical axis for
the curve $\eta=1$ is reduced by a factor of 40.

\bigskip

Fig.4. An example of the unstable spectrum for the transverse
excitations in the state $\eta=1$ under full saturaion for $Q=0$,
$A=2.5\varepsilon _{r}$, $\mu :\mu _{ab}:A=3:1:5$. The scale of
the vertical axis of curve $\eta=-1$ is reduced by a factor of 4.
The cusp in the curve $E_{-}^{2}$, $\eta =1$, corresponds to zero
modulus in (\ref{103}).

\newpage

\begin{figure}[htbp]
\includegraphics[scale=1.0]{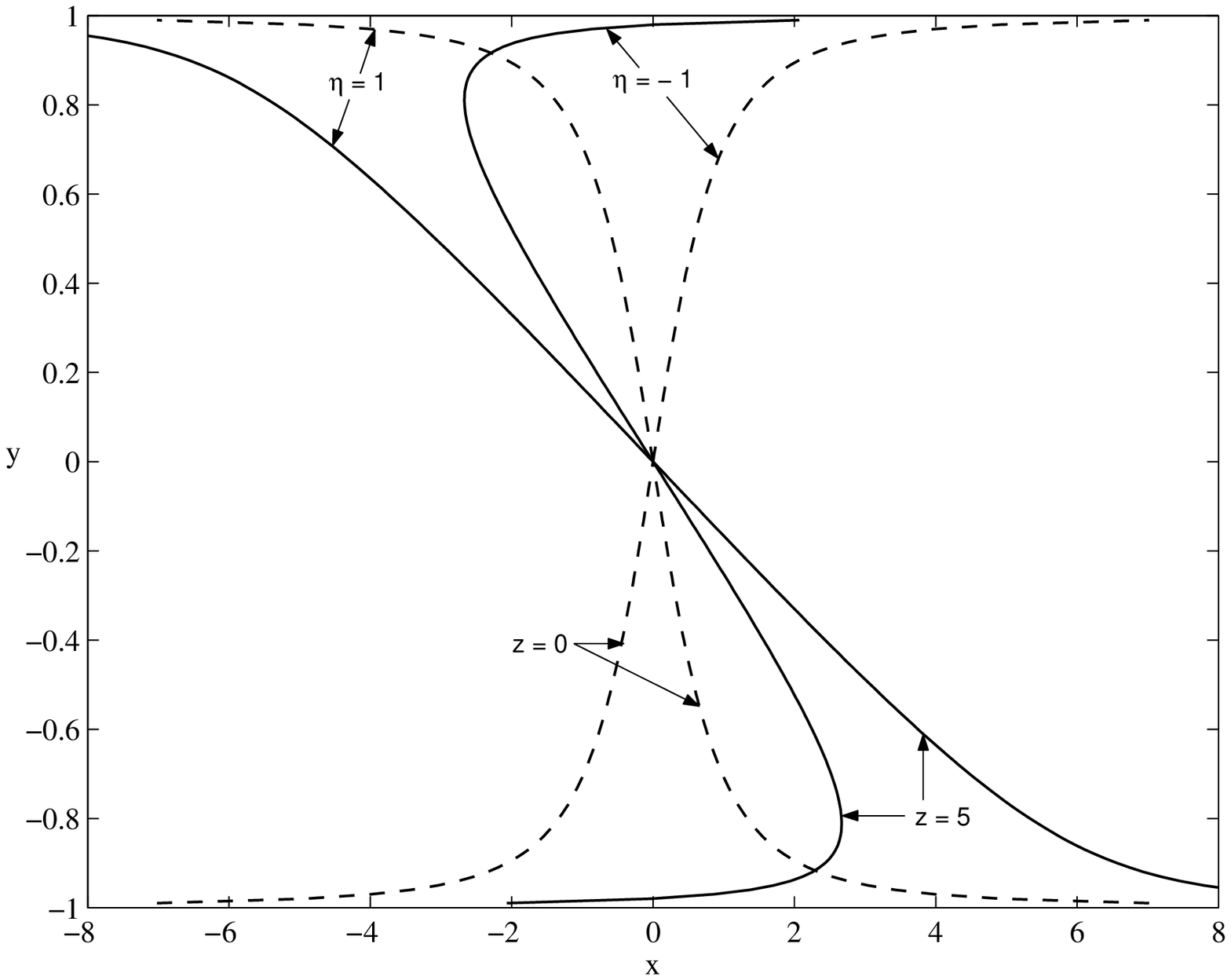}
\caption{}
\end{figure}

\newpage

\begin{figure}[htbp]
\includegraphics[scale=1.0]{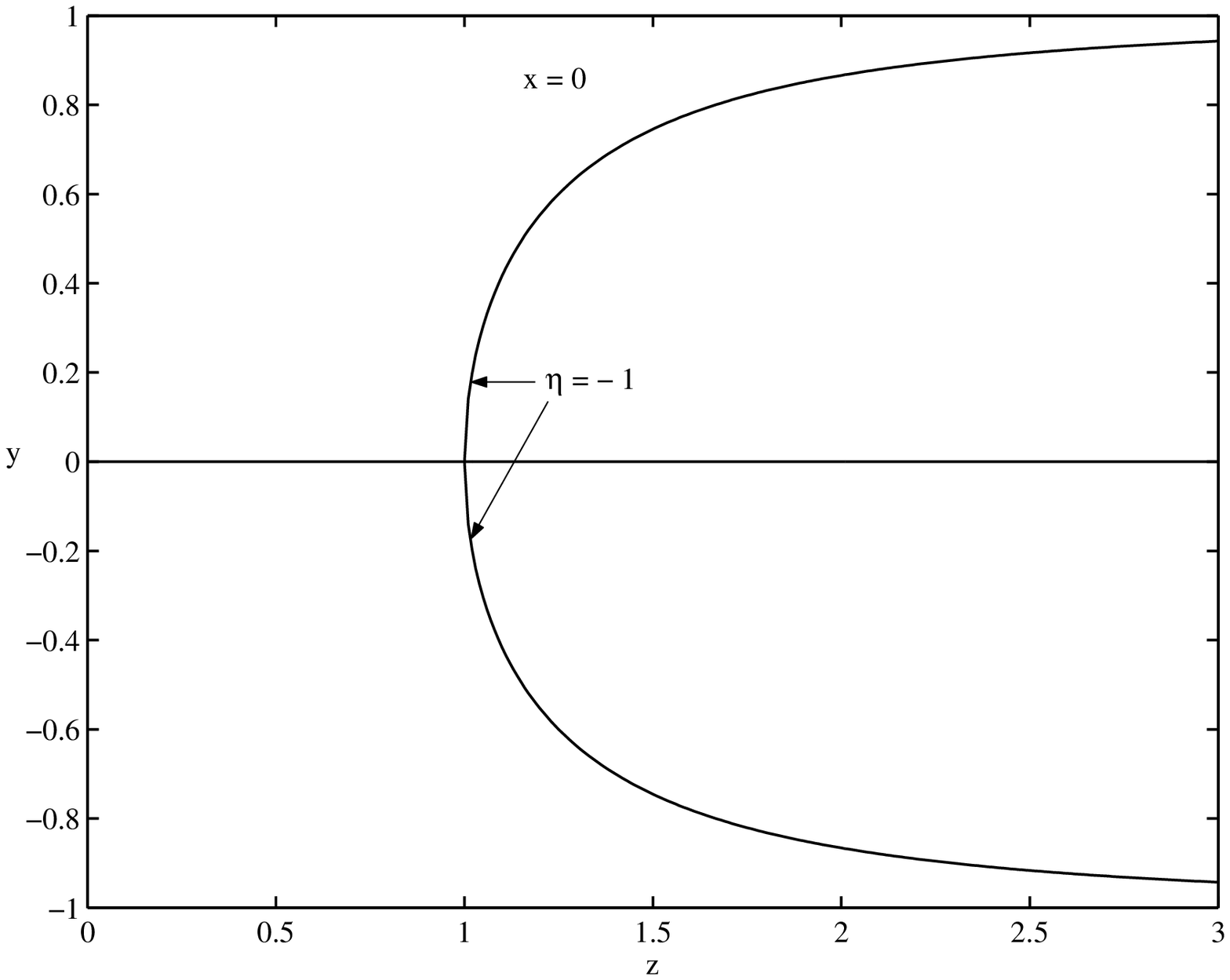}
\caption{}
\end{figure}

\newpage

\begin{figure}[htbp]
\includegraphics[scale=1.0]{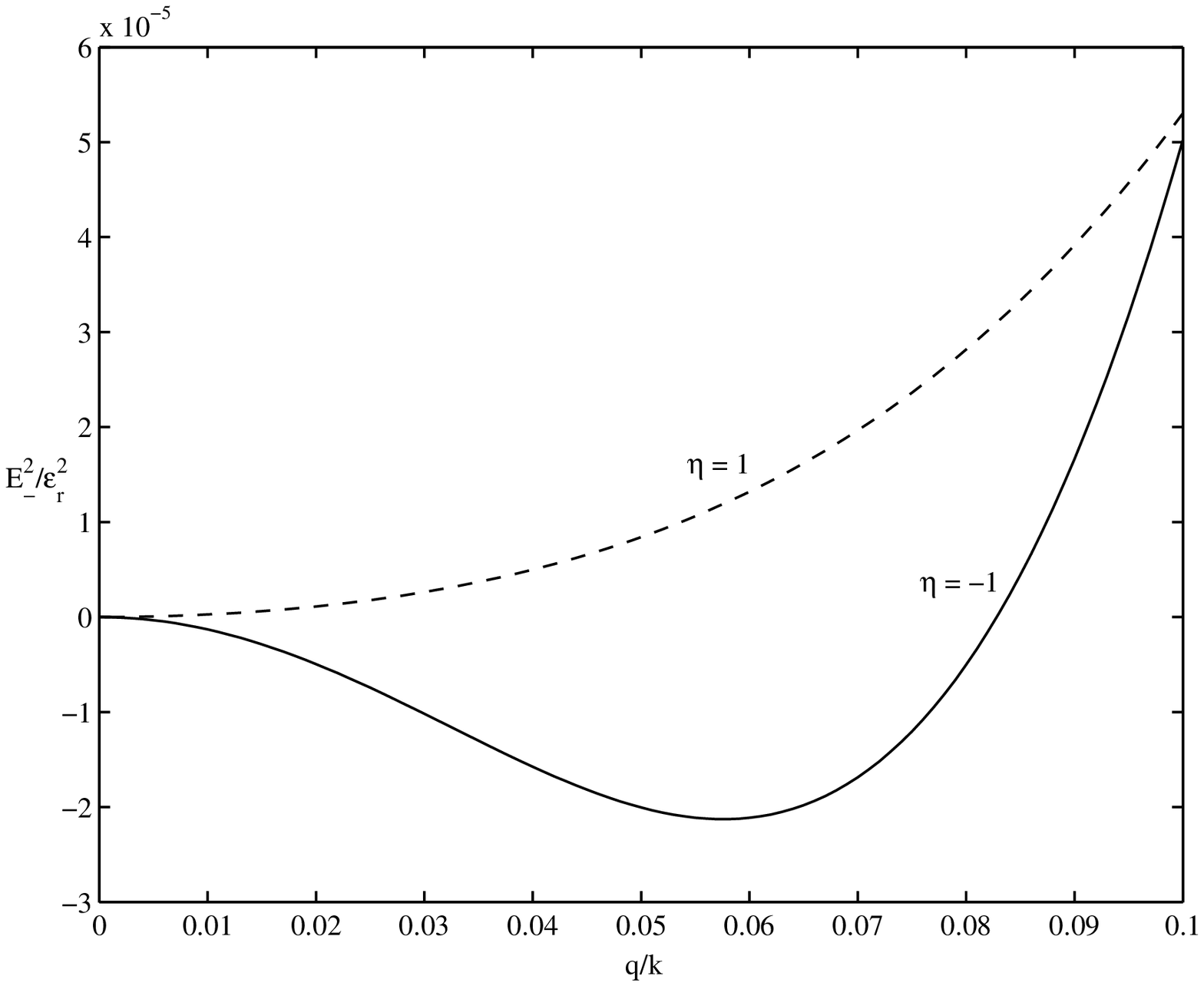}
\caption{}
\end{figure}

\newpage

\begin{figure}[htbp]
\includegraphics[scale=1.0]{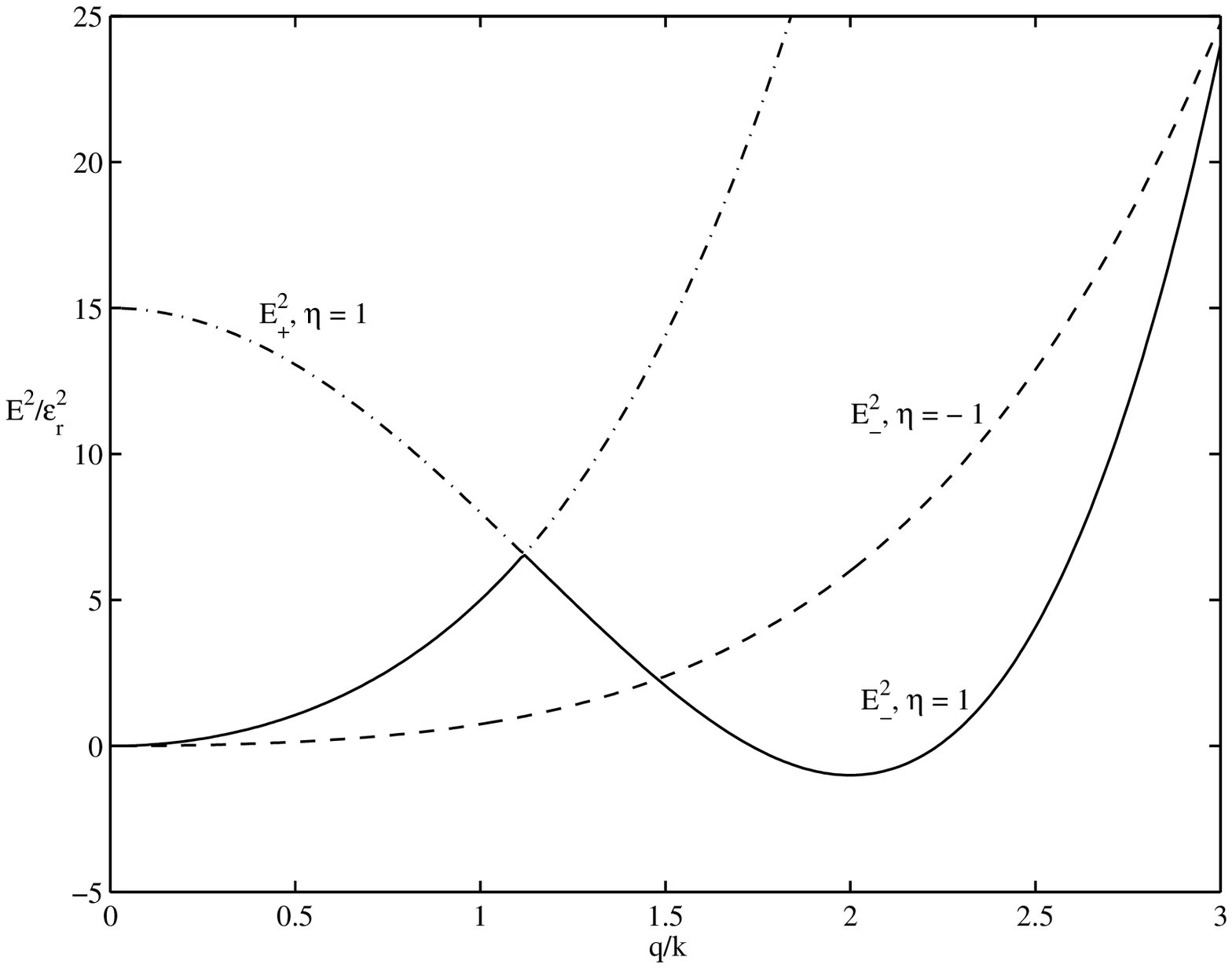}
\caption{}
\end{figure}

\end{document}